\documentclass[letterpaper, 9pt]{sigcomm-alternate}
\usepackage{commath}
\usepackage{subfigure}
\usepackage{balance}
\usepackage{enumitem}

\usepackage{graphicx}
\usepackage{url}

\usepackage{setspace}
\usepackage[hang]{footmisc}

\graphicspath{{figures/}}

\usepackage{etoolbox}
\usepackage{xstring}
\DeclareListParser{\doslashlist}{/}
\newcounter{ndnNameComponentCounter}%
\newcommand{\name}[1]{{%
  \setcounter{ndnNameComponentCounter}{0}%
  \renewcommand{\do}[1]{{%
    \ifnumgreater{\value{ndnNameComponentCounter}}{0}{\allowbreak/}{}%
    \ifnumodd{\value{ndnNameComponentCounter}}{}{}%
    \detokenize{##1}}%
    \stepcounter{ndnNameComponentCounter}}%
``{\fontfamily{cmtt}\small\selectfont\IfBeginWith{#1}{/}{/}{}\doslashlist{#1}}''%
}}

\usepackage{color,soul}
\usepackage{xspace}

\numberofauthors{3}
\author{
\alignauthor
       Spyridon Mastorakis\\
       \affaddr{UCLA}\\
       \email{mastorakis@cs.ucla.edu}\\
\alignauthor
       Kevin Chan\\
       \affaddr{US Army Research Laboratory}\\
       \email{kevin.s.chan.civ@mail.mil}\\
\alignauthor
       Bongjun Ko\\
       \affaddr{IBM T. J. Watson Research Center}\\
       \email{bko@us.ibm.com}\\
\and 
\alignauthor
       Alexander Afanasyev\\
       \affaddr{Florida International University}\\
       \email{aa@cs.fiu.edu}\\
\alignauthor
       Lixia Zhang\\
       \affaddr{UCLA}\\
       \email{lixia@cs.ucla.edu}\\
}

\begin{document}

\title{Experimentation With Fuzzy Interest Forwarding in Named Data Networking}

\maketitle

\begin{abstract}
In the current Named Data Networking implementation, forwarding a data request requires finding an exact match between the prefix of the name carried in the request and a forwarding table entry.
However, consumers may not always know the exact naming, or an exact prefix, of their desired data.
The current approach to this problem---establishing naming conventions and performing name lookup---can be infeasible in highly ad hoc, heterogeneous, and dynamic environments: the same data can be named using different terms or even languages, naming conventions may be minimal if they exist at all, and name lookups can be costly.
In this paper, we present a fuzzy Interest forwarding approach that exploits semantic similarities between the names carried in Interest packets and the names of potentially matching data in CS and entries in FIB.
We describe the fuzzy Interest forwarding approach, outline the semantic understanding function that determines the name matching, and present our simulation study along with extended evaluation results.
\end{abstract}


\let\thefootnote\relax\footnotetext{This is an extended version of a paper presented at the 13th Asian Internet Engineering Conference (AINTEC 2017), Bangkok, Thailand~\cite{chan2017fuzzy}.}


\section{Introduction}

Named Data Networking (NDN)~\cite{zhang2014-ndn,zhang2010-ndn-tr-1,jacobson2009-ccn} is a new networking approach that realizes the Information-Centric networking (ICN) vision by focusing on data retrieval by names rather than on point-to-point data delivery based on addresses.
The foundation of NDN is the shared use of hierarchical and semantically meaningful namespace between application and network layers, enabling applications to request data by names using Interest packets.
NDN routers use names carried in Interests to decide where to forward each Interest if the requested data is not available in the local content store (CS), creating pending state for the forwarded Interest.
When the data is found, routers use these pending states to forward data back to one or more requesters.
Both Interest forwarding and data returning use a prefix match operation: an Interest name finding the longest prefix from the forwarding information base (FIB); a data name finding prefix match(es) from the pending Interest tables (PIT).
Note that applications need to know at least a prefix of the data (e.g., \name{/park/yellow-stone/lost&found/dog/...}) in order to retrieve it.
In cases where consumers and producers do not have a full agreement on the exact terminology (e.g., when instead of \name{dog}, consumers or producers use \name{puppy}, \name{canine}, or \name{hound}), consumers may have problems retrieving data,
even though these names are semantically close and the retrieved information can be useful in finding the lost pet.

In this work, we propose a fuzzy Interest forwarding augmentation to
NDN routers that enables discovery and retrieval of data using approximate knowledge of the data namespace.
In particular, as we explain in Section~\ref{sec:whyff}, such fuzzy forwarding can be beneficial in highly dynamic and heterogeneous environments, e.g., ad hoc encounters of tourists in national parks, where the exact namespace knowledge among communicating parties is either infeasible to obtain or costly to discover.
Semantic-aware fuzzy forwarding can enable NDN routers to forward Interest packets with controlled uncertainty and receive potentially relevant and useful data most of the time.
The returned data can also provide a feedback loop that allows applications to learn about the namespace of available data.



The contribution of this paper is four-fold:
\begin{itemize}[leftmargin=*]
\item Identifying new challenges of name discovery in highly heterogeneous networked environments;
\item Introducing the concept of \emph{fuzzy interest} forwarding;
\item Proposing an initial approach to the solution development; and
\item Conducting a simulation study and presenting extensive evaluation results of this initial approach.
\end{itemize}

We provide an overview of the work related to our proposed approach in the following section and a motivating use case in
Section~\ref{sec:whyff}.
Section~\ref{sec:semantic} describes the overview of fuzzy Interest forwarding, and
then we present the details of the semantic name matching and look-up in NDN contexts in Section~\ref{sec:fuzzy-name-matching-lookup}.
In Section~\ref{sec:simulations}, we present our simulation study to evaluate the performance and tradeoffs of fuzzy Interest forwarding. 
Finally, Section \ref{sec:challenges} discusses some additional challenges in realizing the proposed approach, and Section~\ref{sec:conclusion} concludes our work.

\section{Related Work}
\label{sec:related}

This work proposes an extension to the Named Data Networking forwarding mechanism that leverages results from semantic understanding research from the past and recent work. Below we briefly review related concepts and work related to semantic understanding.

\paragraph{Fuzzy Information Retrieval}
With the emergence of the Semantic Web, there have been efforts to do fuzzy information retrieval based on the data and ontologies \cite{Cross1994fuzzy, deALeite2008fuzzyretrieval, Attia2014multiview} in the context of web searches. 
Given a keyword, Internet search engines comb through the database that has a complete knowledge of the world to look for every potential match.  The search results can then be further confined/reordered based on pre-existing information about the user (e.g., user's location, pre-configured preferences, previous search history, etc.). 
In some approaches, there may be a lack of consistency across similar data across multiple ontologies. While there is a significant amount of work in the area of web search, there is a clear benefit to studying fuzzy information forwarding, as the information matching and networked information flows (both upstream and downstream) are integrated.



\paragraph{Attribute-Based Scheme}
There exists a number of proposals for attribute-based or name-value pair resource discovery schemes.
Adjie-Winoto et al.~\cite{adjie1999design} proposed the Intentional Naming System (INS) and approach that routes information using predefined naming conventions.
The INS resolvers are responsible for information routing as well as updating any additional named data in the network.
To address uncertainty in the name-value pairs, INS allows for wildcard values, but do not exploit any similarities between the resources or their names.
INS assumes a static environment with a given set of attribute names, thus it does not admit any dynamic learning or feedback of the data or namespace.
It is designed as an application overlay, running on top of an existing (IP) network, instead of providing network layer delivery services.



\paragraph{Naming Conventions for Name Discovery}
In NDN, when users or applications know the exact data names, they simply express Interests for this data.
However, for dynamically produced data or data that can be produced by different publishers (i.e., distributed data production), the exact name may not be known a priori.
NDN allows consumers to fetch data using prefixes, which can bring back any data under that prefix.
Based on the received data and following established naming conventions, the user can deduce what (exact) name(s) to use to fetch the next (other) data piece(s). 
For example, to discover current alerts about Yellowstone park, an application can send an Interest for \name{/park/yellowstone/alerts}, which returns a version of the alerts page, which could include additional information about the version, e.g., \name{/park/yellowstone/alerts/_v=2017-04-23}.
With understanding of versioning naming conventions, the application can then request a more recent version or one from several days ago.
Establishing naming conventions can be a powerful tool in name discovery; however it assumes all communicating parties share well established naming conventions a priori.

\paragraph{Metadata for Name Discovery}
The naming conventions provide knowledge about the namespace structure.
In more dynamic environments or where it is impossible to define a standard set of namespace categories, one may consider to use ``metadata'' to describe names that can appear under a given prefix.
For example, a published version of metadata for the park, (e.g., \name{/park/yellowstone/lost&found/_metadata}) can list all report categories for the lost and found items and pets, currently known to the ranger station.
Using metadata for name discovery costs additional round trips, can be expensive when the list of categories is large, and have a limited value when categories could not be aggregated at a central point (e.g., when park visitors do not have connectivity to the ranger station).





\section{Motivation \& Use Case}
\label{sec:whyff}

Our target scenarios are network environments with high degrees of heterogeneity and dynamics, where the namespace of the communicated data can be defined only partially.
As a specific use case in this paper, we use potential NDN-based ad hoc communication between visitors in popular areas of a national park (e.g., searching and reporting missing pets near Old Faithful geyser in Yellowstone national park); we believe similar scenarios can be found in many other applications with high uncertainty and dynamics.

Communication between park visitors are typified by significant dynamics, limited knowledge, and lack of resources. 
Specifically, users and devices' mobility lead to network topologies and link states vary with time. 
Visitors come from different parts of the world, having different understanding of terminology to describe data, contributing to heterogeneity of data naming and naming convention.
There may also be limitations in resources and limited (or no) access to the infrastructure-assisted information retrieval.
For example, reports about lost pets or sighting of wild animals may not be able reach a ranger station for information aggregation, but would be immediately useful to visitors in the area.

In this paper, we consider non-malicious users trying to publish animal sighting reports and users who want to retrieve information, based on some partial knowledge of the reports' name.
We assume that there is a large report amount available in the network and limited communication budget, making infeasible to retrieve the entire dataset or even enumerate the namespace.
We assume that users (applications running on their devices) have some namespace understanding, e.g., they would know that all lost\&found reports in the park are published under the \name{/park/yellow-stone/lost&found} prefix and all wild animals sightings near Old Faithful under \name{/park/yellowstone/old-faithful/alerts}.
%
To retrieve data, users construct Interests under the specific prefixes, adding potentially ambiguous parts to describe the desired information (\name{dog}/\name{pup} when searching for a lost pet, \name{bison}/\name{buffalo} or \name{bear}/\name{grizzly} when looking for wild animal alerts).
The constructed Interests may also include suffixes to describe the requested information type, such as \name{info} or \name{video} for a text description or a video feed, and other meta parts (e.g., version and segment numbers).
Users then express such Interests to the NDN network that includes their ad hoc peers or any ranger stations in sight, which utilizes fuzzy interest forwarding and data matching to retrieve data semantically close to the request, but not necessarily has the same exact prefix.
After receiving initial reply, users can refine their question and retrieve other associated data, e.g., other segments of video report.

\section{Fuzzy Interest Forwarding (FIF)}
\label{sec:semantic}

\begin{figure}[t!]
  \centering
  \includegraphics[scale=0.5]{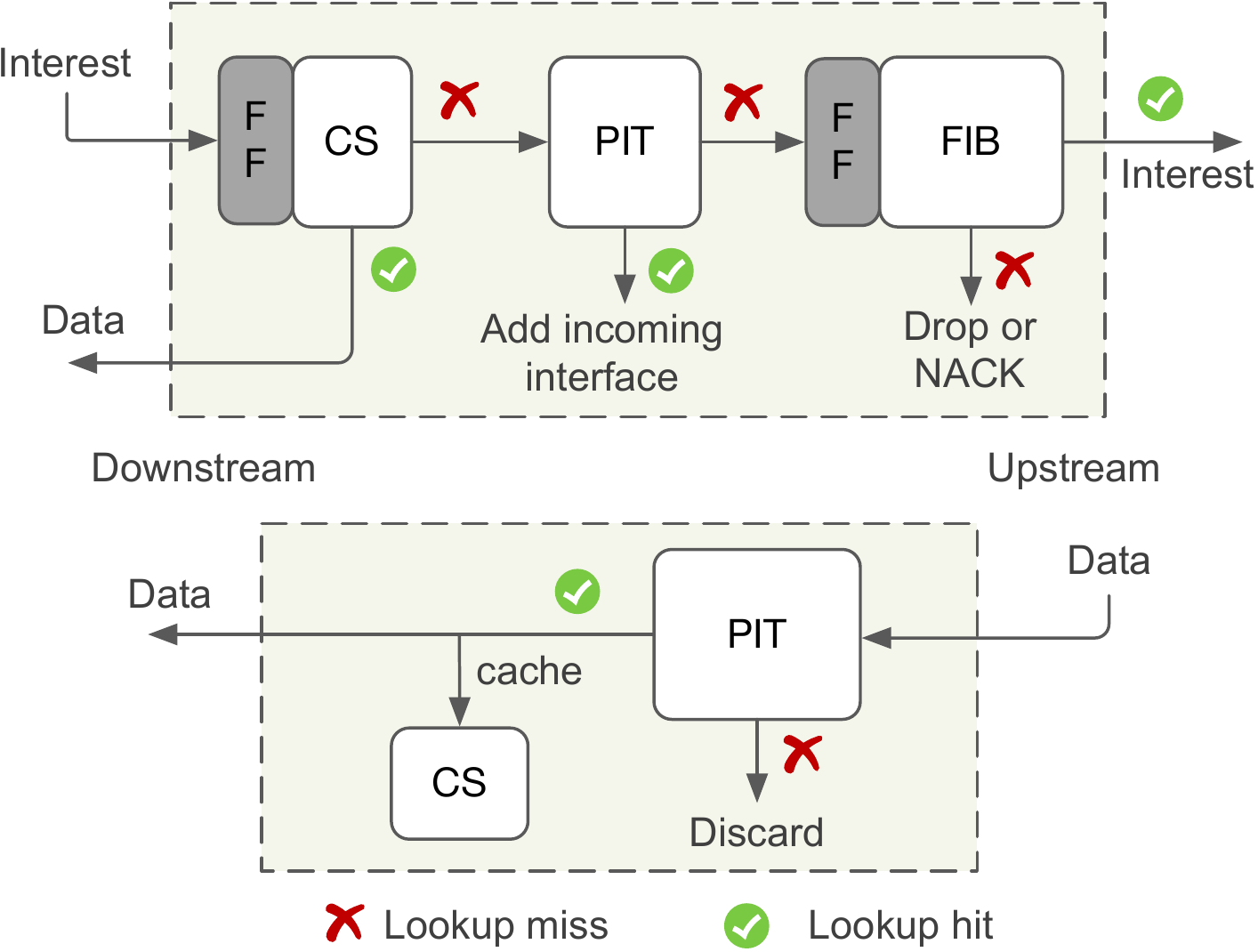}
  \caption{NDN forwarding strategies with fuzzy forwarding}
  \label{fig:router}
\end{figure}

This section describes the architecture of augmenting NDN with Fuzzy Interest Forwarding (FIF). 
As highlighted in Figure~\ref{fig:router} with grayed ``FF'' blocks, we propose to add fuzzy matching operations at two stages as an Interest packet goes through an NDN forwarder: the Content Store (CS) and the Forwarding interest base (FIB).
To match the retrieved Data with the Interest in the PIT, we plan to rely on a recently proposed \emph{interest digest} mechanism, highlighted later in this section.

\subsection{Augmented Interest Forwarding}

Overall, the fuzzy Interest forwarding mechanism shown in Figure~\ref{fig:router} is similar to the standard Interest processing by an NDN router~\cite{zhang2014-ndn,NFD-guide,yi2013stateful}:
returning data if there is a CS match, creating a new PIT entry or aggregating with an existing one, forwarding the Interest based on a matched FIB entry.\footnote{The actual forwarding decision is made by the forwarding strategy, which we omit for simplicity.}
However, matching in the CS and FIB is performed not just based on the prefix match, but also considers the semantic closeness between some of the name components.
In other words, when looking up the CS and FIB, an NDN router uses a part of the Interest name as usual to select a set of potential candidates, then uses one or more name components to calculate fuzzy semantic distance with the candidates, potentially following with further matching on the name suffix (exact or fuzzy).
We leave to the future research the decision of how exactly the router determines which part of the name should be matched exactly and which fuzzy.
One potential direction is to designate a dedicated name component type.




When an Interest fails to find a fuzzy match in CS (i.e., there are no data matching the non-fuzzy prefix or the fuzzy match resulted in a low-confidence match, see Section~\ref{sec:fuzzy-name-matching-lookup}), it is fuzzy matched with entries in the FIB.
For such fuzzy-forwarded Interests, the returned Data most likely will not share the same prefix.
One way to address this issue is to use a recently proposed \textit{interest digest} mechanism~\cite{ndn-team2016interest-digest} to uniquely identify the Interest from the PIT that fetched the returned Data packet.
Whenever a node $N_1$ finds a (fuzzy) matching data for an Interest it received from a neighbor node $N_2$, $N_1$ returns the data to $N_2$ together with the digest (i.e., a cryptographic hash) of the Interest, so that $N_2$ can uniquely identify the Interest in its PIT that fetched the returned data packet.
This way, the Data forwarding path does not require any additional changes to support FIF.


%


FIF also does not require changes to the match between Interests in PIT: Interests carrying different names will create separate entries in PIT, regardless of their semantic distance.
First, it maintains the flow balance because each Interest retrieves at most one (potentially same) data packet.
Second, keeping the original PIT matching logic prevents fuzzy matching of already fuzzy matched names, which may result in a semantic similarity dilution of results returned to the original Interest.

\subsection{Design Considerations}

\subsubsection{Semantic-Based Cache Replacement}

As nodes are generally storage-limited, the CS generally apply a content replacement policy, such as least recently used (LRU), first-in-first-out (FIFO), or others.
CS can also utilize fuzzy matching in the replacement strategy, determining which data should be stored based on a given semantic similarity metric.
For example, CS may prioritize removal of data packets semantically close to each other, keeping only one of them.
This way, CS can conserve its limited storage space, yet be able to return (hopefully useful) data to different incoming Interests. 

\subsubsection{Geo-Based FIB}

To perform a meaningful fuzzy match in FIB, it needs to be filled with entries related to data.
We assume that producers or repositories announce prefixes, which are propagated through the network via routing protocols, such as NLSR~\cite{Lehman16:NLSR,Hoque13:NLSR}.
In the ad hoc context, where usually there is a limited number of interfaces per node, it is possible to use a geo-based FIB (gFIB)~\cite{grassi2015navigo}.
gFIB is still populated through announcements (proactive) or data discovery interests (reactive) from data carriers (producers, publishers, or data mules).
However, instead of designating a ``next-hop'' interface, gFIB entries are associated with geographical coordinates or directions (angles).
Depending on from which direction an Interest has been received, a gFIB entry may tell to forward (re-broadcast) or drop the Interest.

\subsubsection{Use of Fuzzy Match}

Depending on the network and information context, the fuzzy matching described in Section~\ref{sec:fuzzy-name-matching-lookup} can be applied in different ways.
The most appropriate approach is an open question and we leave it for our future work.
In the CS context, fuzzy matching needs to find at most one entry for each Interest, as only one Data can be returned.
Depending on the similarity threshold---another open question and an important design choice---a match can either always return a result, provided there is data that matches the non-fuzzy Interest part, or return when a ``reasonably'' close match is found.

In the FIB context, the fuzzy match can return zero or multiple entries, found with or without defined semantic similarity bounds.
These entries can be used either
(1)~to forward the Interest based on the next-hop information from ``all'' entries,
(2)~based on top-$k$ FIB entries, or
(3)~to forward based on the ``best'' entry in terms of the distance measure.
In cases when an Interest fails to find any match in FIB, the Interest can be either dropped, or forwarded to a default place where more advanced semantic matching can be performed.

\subsubsection{Forward or Wait Challenge}

The incoming Interest can be semantically matched with different (set of) entries, depending on the current CS and FIB state that changes with each new data or routing announcement.
This poses a challenge for the router whether for each received Interest to immediately apply the fuzzy logic, or delay it by some time in the hope that it will receive more relevant data (requested by previously sent Interests) or prefix announcements.
The more router waits, the more ``precise'' fuzzy decision it can make, while diminishing the requester satisfaction. 
Specific to the fuzzy name matching, we can also envision user feedback to refine the returned fuzzy matched data (further discussion in Section~\ref{sec:challenges}).


\section{Fuzzy Name Matching and Lookup}
\label{sec:fuzzy-name-matching-lookup}

In this section, we discuss in detail the issues of matching the names\footnote{For simplicity, we talk as if fuzzy match is performed on the whole name.  As we described in Section~\ref{sec:semantic}, the actual fuzzy match may include only certain components, either designated by the consumer or producer.} in NDN by their semantic similarity and how to perform the look-up of the semantically matching names in FIB and CS.

\subsection{Fuzzy Name Matching}
\label{sec:fuzzynamematching}

With fuzzy name matching, a name is matched to other name(s) in CS and FIB based on a certain measure of their semantic  similarity.\footnote{
In this paper we do not consider the syntactic similarity, which can be efficiently handled by existing methods~\cite{navarro2001guided}.
}
To formally define such a similarity measure, let us denote by $D(n,n')$ a distance function between two names $n$ and $n'$, where a small $D(n,n')$ means $n$ and $n'$ are similar ($0$ for the same or ``equivalent'' names, close to $\pm1$ for very different names). Then the fuzzy matching of a given name $n$ (of an Interest) to a set of names $S$ (in CS or FIB) can be defined as either

\begin{itemize}[leftmargin=*]

\item finding a name $s$ in $S$ that minimizes $D(n,s)$ ({\it Best-Match}), or

\item finding names in $S$ such that $D(n,s) < \tau$ for some threshold constant $\tau > 0$ ({\it Qualified-Matches}).
\end{itemize}

In practice, we expect the combination of the above matching methods to be most useful, which returns the best-matching name only if its distance to the given name is within a certain threshold. In what follows, we review two semantic distance measure categories and methods to obtain them in NDN.
We first consider case when only one name component with a single word is matched against another single word.
We discuss more complex fuzzy matching cases on multiple name components in Section~\ref{fuzzy-matching-hierarchical}.

\subsubsection{Ontological matching}

\begin{figure}
  \centering
  \includegraphics[scale=0.45]{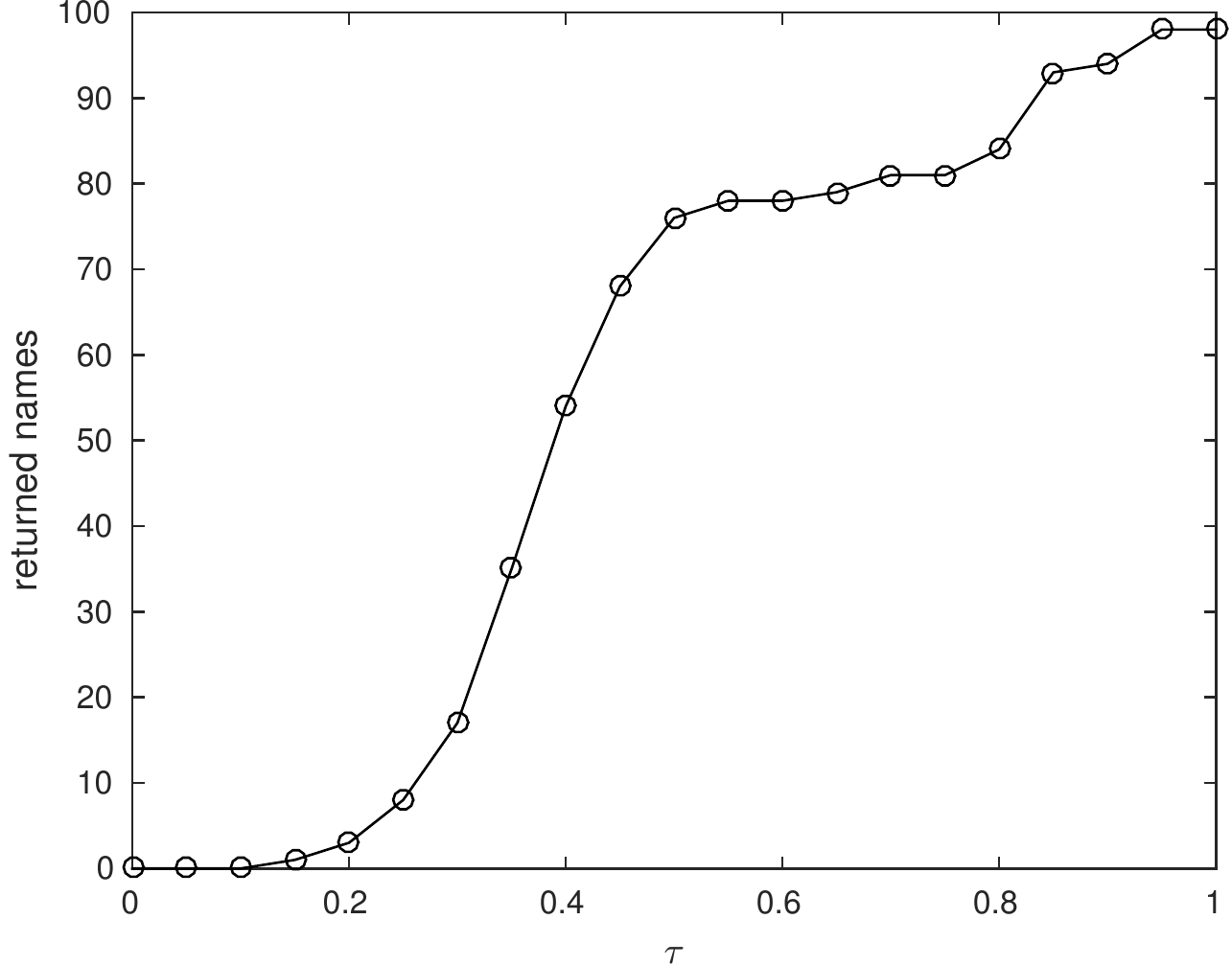}
  \vspace{-3mm}
  \caption{Names vs $\tau$ using WUP over the Wordnet Synset corpora matching wildebeast to the UCI Zoo dataset.}
  \label{fig:names_vs_tau}
\end{figure}

With ontological matching, the similarity between two names is found based on a pre-defined name ontology given by a lexical database such as WordNet~\cite{miller1995wordnet}. A variety of the semantic distance measures have been proposed and used in the literature, but essentially they are typically defined by some function of the topological distance between the nodes in the ontological tree.

In Figure~\ref{fig:names_vs_tau}, we present the result of an experiment to demonstrate how we envision FIF would be affected by the choice of the threshold parameter $\tau$ in the ontological distance measure. We use an example from the UCI Machine Learning Repository Zoo Data Set~\cite{Lichman:2013} as the input data, the Synset in WordNet corpus as our ontology, and the WUP Similarity score by Wu and Palmer~\cite{wu1994verbs} as the semantic similarity function. The figure shows an (obvious) increase in the number of similar names being returned as the threshold $\tau$ increases. We also see a sharp transition at around $\tau = 0.3$, which indicates the parameter choice plays an important role as a control knob to tradeoff the accuracy with the overhead of fuzzy forwarding.

Since the ontology-based matching uses a pre-defined dictionary curated by people over time, it can provide a well-defined set of semantically similar words for a word majority used by humans. The major drawback is that a large database needs to be loaded in NDN routers (consider NDN forwarder running on users' phones in our example) and queried each time a name is to be matched against others, which would prohibitively increase the cost both in terms of memory and processing time. For this reason, we envision this method would be appropriate when the fuzzy name matching can be performed in resource-rich computing nodes, e.g., at the ingress point of the network or server-scale middle boxes to which the Interest processing is delegated.

\subsubsection{Contextual matching}

\begin{figure}
  \centering
  \includegraphics[scale=0.23]{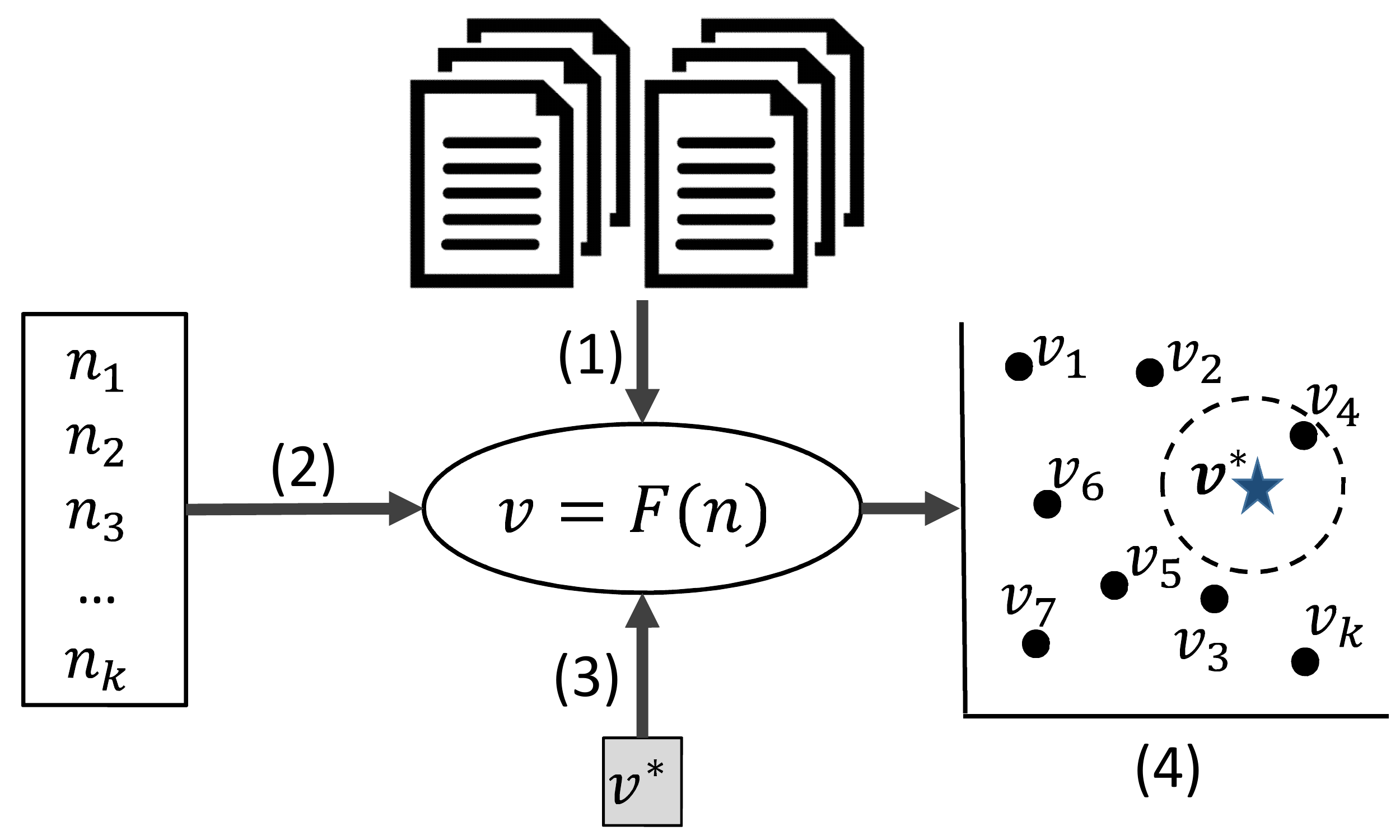}
  \vspace{-3mm}
  \caption{Constructing and using vector space model for contextual similarity distance between the words. 
{\small (1) A name-to-vector mapping function is learned from a corpus of texts, (2) names in CS and FIB are mapped to vectors, (3) a new name (query) is mapped to a vector, and (4) semantically similar names are found in the vector space through some distance measure between the vectors.}}
  \label{fig:vector-space-model}
  \vspace{-3mm}
\end{figure}

At a high level, the contextual matching of words is done through the semantic distances learned statistically from a corpus of texts (documents, sentences, phrases) that they are used in. It operates under the assumption of so-called distributional semantic hypothesis, that is, words close in meaning tend to occur in the same piece of texts (same document or same sentence). While there are many different methods for learning the contexts, they generally work by learning a mapping function that converts the words into fixed-length vectors in some latent vector space of much smaller dimension than the original corpus of words. Then the distance between the mapped vectors represents the semantic distance between the words (See Figure~\ref{fig:vector-space-model}). 

There are two broad classes of mapping functions and how they are learned: context-counting methods and context-predicting ones.
The former (more classical) approaches, represented by Latent Semantic Analysis (LSA)~\cite{landauer2006latent}, discover the latent features of the words based on their co-occurrence counts in the set of documents.
Whe latter (more recent) breed focuses on the ``prediction context'' that indicates a word's influence on predicting other surrounding words in the text~\cite{le2014distributed}. Details on the difference and similarity between these two approaches can be found in literature~\cite{baroni2014don}.

%

With these approaches, our distance function $D(n,n')$ would be defined as $D(n,n')$=$d(v, v')$, where $v=F(n)$ and $v'=F(n')$ are the respective n-dimensional vector representations of names $n$ and $n'$ determined by the name-to-vector mapping function $F$, and $d(\cdot)$ is a distance function in the vector space. A typical distance measure is the cosine distance:

\begin{equation} \label{cosine-dist}
	d(v, w) = 1 - \frac{v \cdot w}{\norm{v} \norm{w}},
\end{equation}

\noindent where $v \cdot w$ is a dot product of two vectors $v$ and $w$, and $\norm{v}$ and $\norm{w}$ their magnitudes.

These approaches have distinct advantages over the ontology-based approaches. First, the semantic relationship between the names is represented by a compact vector model, hence not requiring the entire database of the names and their relationship to be loaded in memory for fuzzy matching. Also, these methods can inherently provide the semantically similar names within the context of how the names of data are being requested. Finally, the distance calculation between the words are done through simple algebraic operation in vector space, making them adequate for resource-constrained NDN routers. Due to these benefits, in what follows we further explore some details of how the contextual matching can be effectively utilized in NDN environments.

\subsection{Constructing Vector Space Model of Names}

A potential issue of using the vector space model for contextual name matching is that its applicability and effectiveness depends on the dataset used to train and build the model since a similarity between two names can be found only when they have been seen in the training dataset. Here we present a few alternatives and their pros and cons when used in NDN contexts.

One approach is to build the vector model with the NDN names observed (published and used). The name contexts can be obtained from the hierarchical name constructs (prefixes and suffixes) as the ``sentences'' in which the individual words are used. The challenge, however, is that names only can provide the context to only a limited extent so the accuracy of the model may be low. Also, because the model is built and evolved dynamically as new observations are made, it needs to be somehow synchronized dynamically across all routers.\footnote{It should be possible to have each NDN router build its own model based on what it observes, though it implies additional challenges regarding the model being more sparse and the potential network-wide instability of the fuzzy forwarding due to heterogeneous forwarding decisions.}

Another approach at the other end of spectrum is to build a model from a large corpus of public data set, such as Wikipedia documents, that includes a rather comprehensive set of words and their contexts. With this approach, a pre-built, static model with relatively good coverage can be used across all NDN routers. The downside, however, is that the vector space model itself can become very large, and there is still a chance that some names may not be present in the model.

A hybrid of the above two approaches can potentially strike a good balance, which uses corpora of online documents and other meta information on the data themselves that exist in the network in addition to the associated names, in order to enrich the contextualization of the vector space model. How to synchronize the learned model across the network still remains an open problem, to which a variety of solution strategies can be considered--from a centralized control plane, to an NDN-approach to model distribution (by assigning the models unique names), to a decentralized (possibly localized) model synchronization between nearby nodes.

\subsection{Fuzzy Name Look-up in Vector Space}

The final step of fuzzy forwarding is to find, among all names in FIB and CS, the nearby ones to a given name in the vector space (i.e., step (4) in Figure~\ref{fig:vector-space-model}). Whether the goal is to find the most similar one or those within a certain distance threshold, this step involves searching for nearest neighbor(s) in the vector space. A na\"{\i}ve method for the nearest neighbor search would evaluate the pair-wise distances between the query vector and all vectors, resulting in $O(n)$ search cost, which could take prohibitively long in resource-constrained routers. To address that, we suggest the use of approximate nearest neighbor (ANN) search methods such as Locality Sensitive Hashing (LSH)~\cite{slaney2008locality}, which provide $O(1)$ or $O(\log n)$ look up time at the small expense of additional memory requirements for indexing data points.

\begin{figure}
     \centering
     \subfigure[Comparison of query response time] {
        \label{fig:lsh-time}
        \includegraphics[scale=0.45]{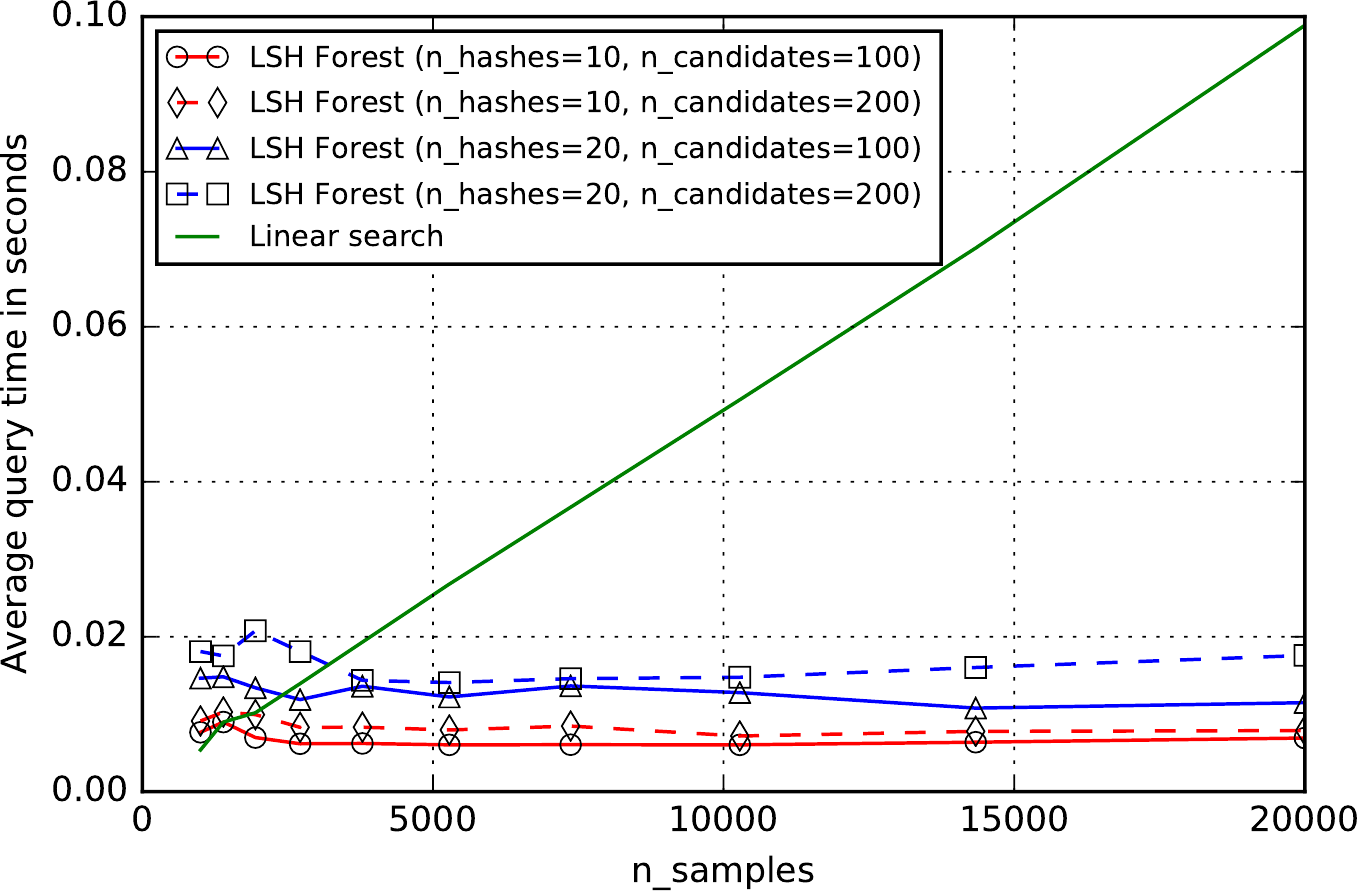}}
    \subfigure[Accuracy of LSH Forest method for nearest neighbor search] {
        \label{fig:lsh-accuracy}
        \includegraphics[scale=0.45]{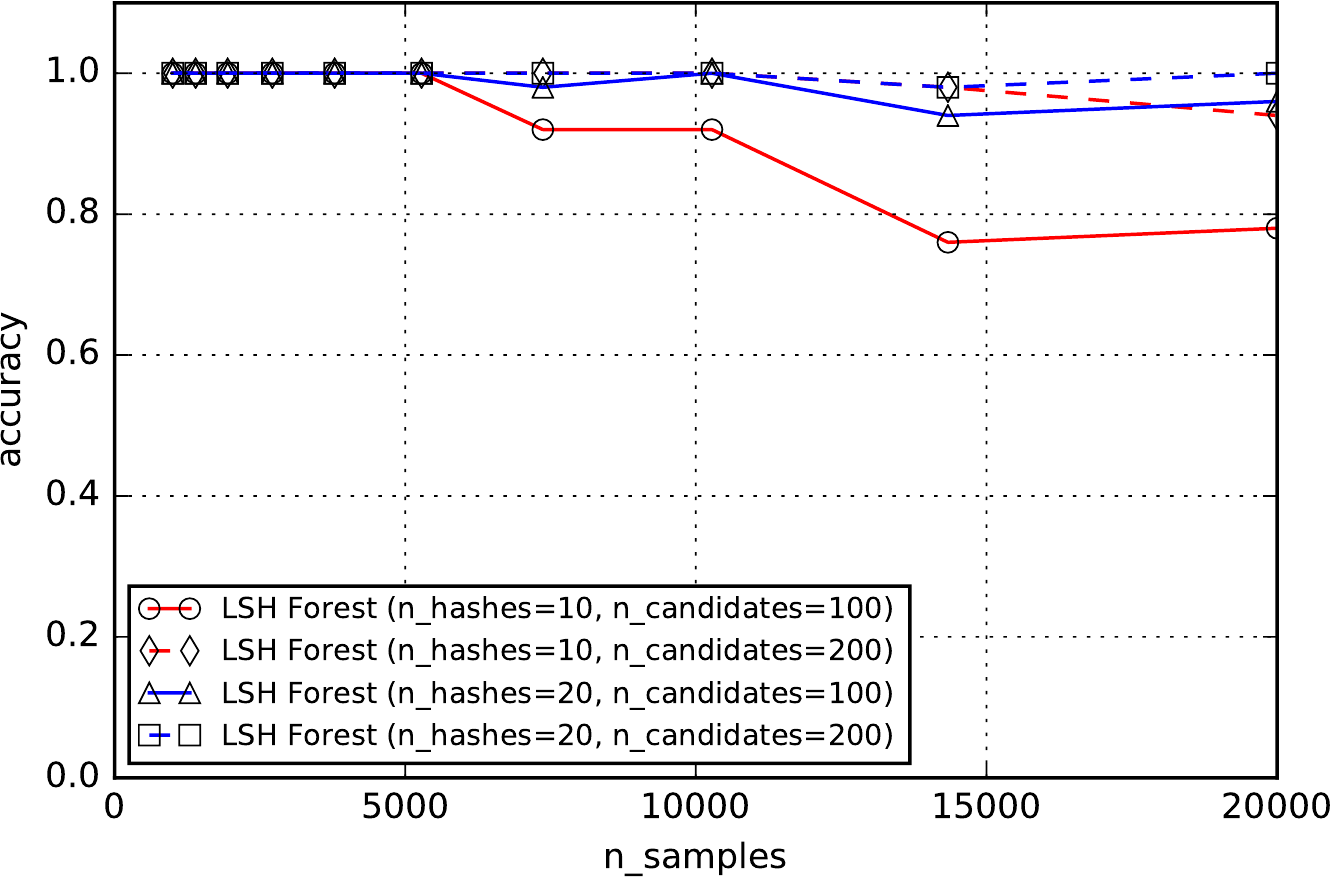}}
     \vspace{-3mm}
    \caption{Performance of ANN search using LSH Forest algorithm.}
    \label{fig:lsh-performance}
    \vspace{-3mm}
\end{figure}

As a proof of concept, we conduct a small experiment to measure the performance of an ANN search using LSH Forest~\cite{bawa2005lsh} method for 400-dimensional vector space. In Figure~\ref{fig:lsh-time}, we compare the ANN query response times of LSH Forest method with different parameters against a brute-force linear search as we vary in the $x$-axis the total number of data points to which the given query vector is compared against. Figure~\ref{fig:lsh-accuracy} shows the average accuracy of the ANN search result returned by LSH Forest method. The results indicate that LSH Forest method (or other ANN methods) produce reasonably accurate NN search results with almost constant look-up time, making them viable choices for fuzzy name search in resource-constrained NDN routers.

In addition, it would be beneficial to quickly determine whether a name semantically matched to the given one {\it exists} in a set of names (e.g., CS or FIB for individual outgoing interface) so that the actual look-up is performed only when there is a match. For such a membership query, the use of Bloom filter has been proposed in literature for exact name query in NDN~\cite{quan2014scalable}. In the similar spirit, we could explore the use of locality-sensitive Bloom filters (LSBF)~\cite{hua2012locality} for the {\it approximate} membership query in the vector space model of the names.

\subsection{Multi-Component Fuzzy Match}
\label{fuzzy-matching-hierarchical}

NDN names are shared between the application and the network, therefore, as we highlighed in Section~\ref{sec:semantic}, some parts of the name still need to be exactly matched.
For example, \name{/park/yellowstone/lost&found} should always be matched exactly (i.e., non-fuzzy) to avoid retrieval of irrelevant data from other parks or wild animal alerts instead of the lost\&found reports.
However, how many components of the name can be fuzzy-matched (who and how they are designated) is an open question.
Here we discuss two approaches to realize fuzzy matching that considers multiple components of an NDN name, having their own advantages and limitations that we plan to explore in our future research.

One approach to realize the multi-component fuzzy match is to make it a superposition of the word-by-word (component-by-component) fuzzy matches at the corresponding positions in the sequences.
Specifically, given two multi-component sequences, the measure of their semantic distance could be some combination of the component-wise distances at individual positions.
However, the name hierarchy itself can be different from one namespace to another (e.g., \name{../animal/dog/..} versus \name{../animal/mammal/canine/hound/..}).
Therefore, it may not be sufficient to compare the words at the same level only, and the distance measure may need to consider all cross-level distances, possibly assigning higher weights to the same-level comparisons to account for a higher relevance in the overall assessment of the similarity.

Another approach is to compare multi-component sequences {\it as if} they are flat ones.
In other words, one can build a vector space model, not for individual words, but for the entire sequences by treating them as phrases or sentences---in much the same way used for topic modeling and sentiment analysis of texts.
One then would compare the entire ``sentences'' in the vector space to find their similarity; e.g., the ``sentence'' of \name{../pet/puppy/hound/..} would be mapped to a vector close to one mapped from \name{../animal/canine/dog/..}.

\section {Simulation Study}
\label{sec:simulations}

In this section, we present our simulation study including extensive evaluation results. Our goal through this study is to evaluate: 1) The performance of FIF compared to Exact Matching Forwarding (EMF), 2) The overhead imposed by FIF, 3) The scalability of FIF in terms of the number of matches and the semantic similarity threshold, 4) The performance of the ``Forward or Wait'' approach, and 5) How ``similar''  the retrieved data is to the data requested by the consumer application (i.e., the retrieval accuracy as observed by the application).

\subsection {Experimental Setup}
\label{setup}

\begin{figure}[t!]
  \centering
  \includegraphics[scale=0.32]{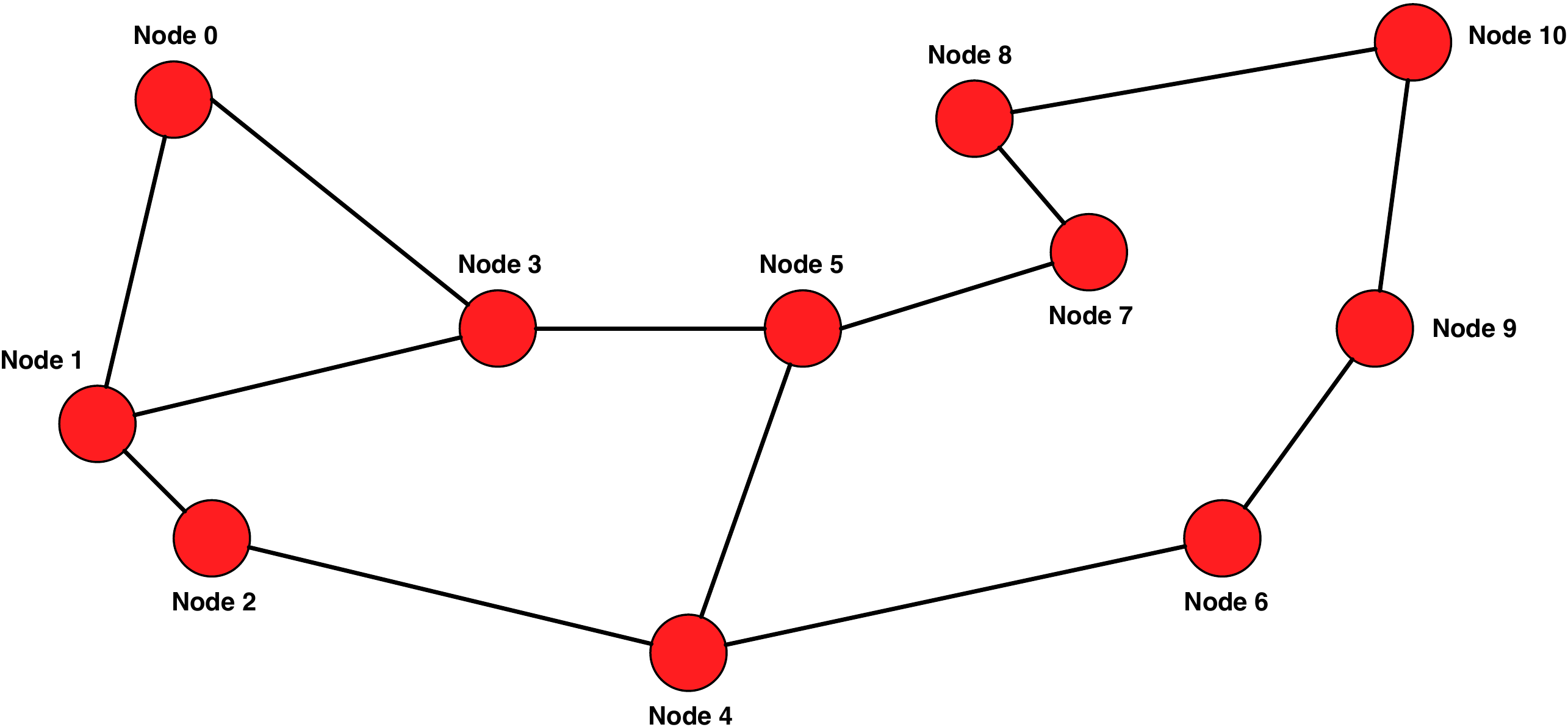}
  \caption{Simulation Topology}
  \label{fig:topo}
\end{figure}

For our simulation study, we implemented FIF in ndnSIM~\cite{mastorakis2017evolution}\footnote{The implementation is available at \url{https://github.com/spirosmastorakis/FIF}}, the NS-3 based NDN simulator, and we use a topology consisting of 11 nodes and 15 links (Figure~\ref{fig:topo}) with same node distribution and interconnection as the Abilene topology~\cite{Abilene}. The propagation delay between 2 topology nodes is 10ms.

Our implementation performs single name component fuzzy matching through Google's word2vec tool~\cite{word2vec} (we use a pre-trained model consisting of 71K words). For the performance evaluation of the ``Forward or Wait'' approach, we deploy one consumer application at node 0 and one at node 3, and a producer application at node 10. Each consumer expresses 20 Interests per second with a separate prefix followed by a fuzzy name component. For all the other simulation scenarios, we deploy a consumer application at nodes 0 and 4, while the producer is still at node 10. Each consumer expresses 10 Interests per second with a common prefix followed by a fuzzy name component. The fuzzy name components are randomly selected from the set of words used to train the used model, since pre-trained word2vec models cannot handle out of dictionary words. 

\begin{figure}[t!]
  \centering
  \includegraphics[width=\columnwidth]{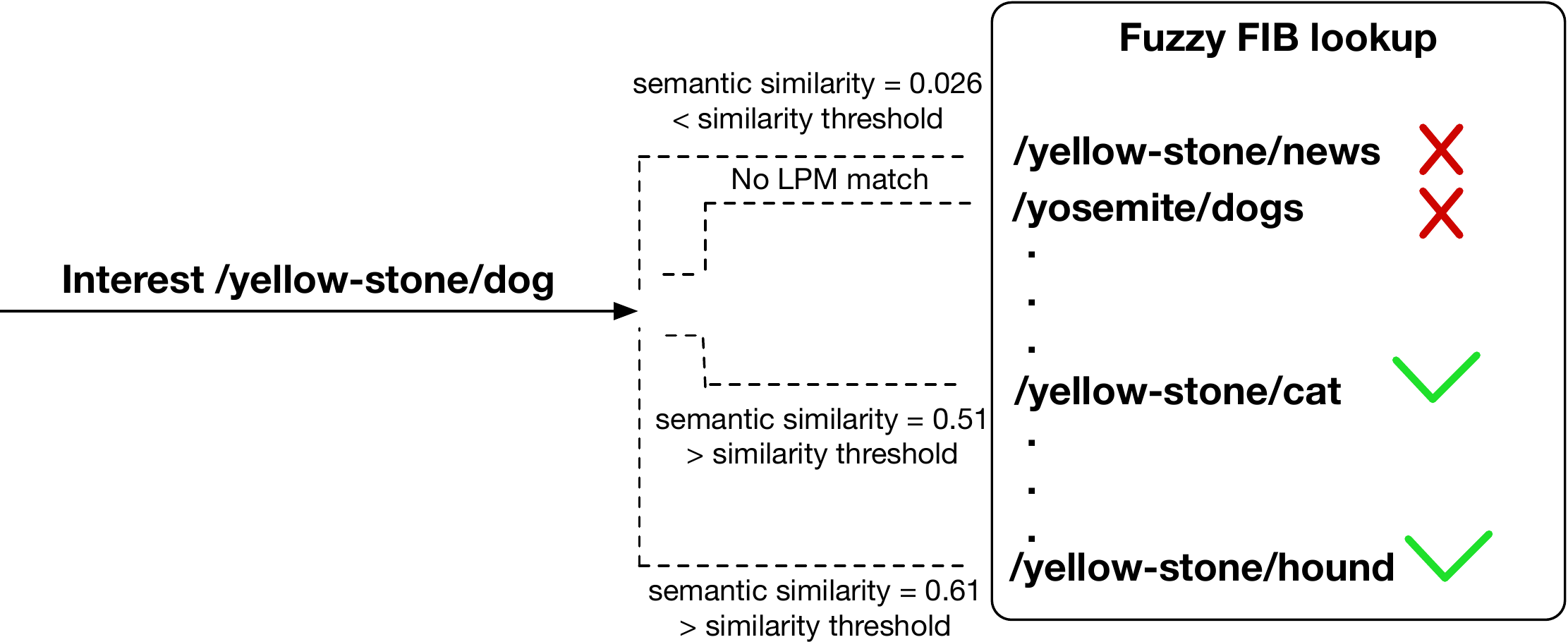}
  \caption{Fuzzy FIB lookup example, semantic similarity threshold = 0.4}
  \label{fig:fib-lookup}
\end{figure}

For each experiment, we specified a maximum number of CS and FIB matches to be performed. In Figure~\ref{fig:fib-lookup}, we present an example of the fuzzy FIB lookup process. For the name prefix of the first FIB entry, there will be no fuzzy match due to a lower cosine similarity than the defined similarity threshold, while for the second one, there is no Longest Prefix Match (LPM) between the Interest prefix and the prefix of the FIB entry. For the FIB entry with prefix \name{/yellowstone/cat}, its similarity with the Interest name is 0.51, which satisfies the specified similarity threshold. At that point, we either have the option to directly forward the Interest based on the matched FIB entry or continue the fuzzy lookup until the specified maximum number of matches is reached. In the case of the latter option, a better match will be found for the FIB entry with prefix \name{/yellowstone/hound} and cosine similarity of 0.61 to the Interest name. The fuzzy CS lookup process is similar to the presented fuzzy FIB lookup process.

We ran each simulation scenario 5 times, while each run lasted 70 seconds. During the first 10 seconds of the simulation, we populated the CS of each node with different data to create node heterogeneity, while the actual data retrieval of the consumers from the producer began afterwards and lasted 60 seconds.

\subsection {FIF Data Retrieval Performance}

\begin{figure}[t!]
  \centering
  \includegraphics[width=\columnwidth]{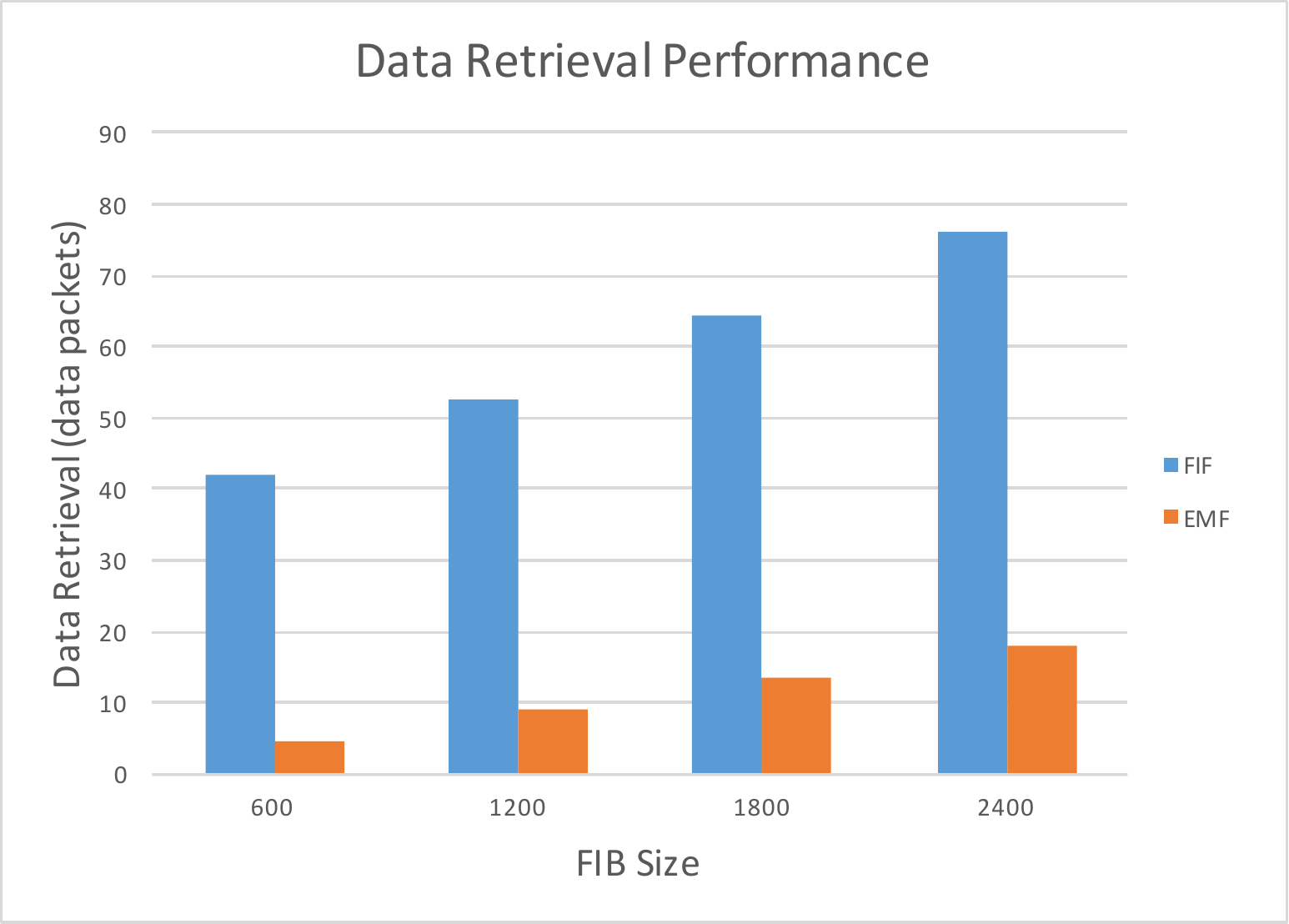}
  \caption{FIF and EMF performance, number of matches = 50, semantic similarity threshold = 0.4}
  \label{fig:perf}
\end{figure}

In Figure~\ref{fig:perf}, we present the performance of FIF and EMF as we vary the FIB size. The results  show that the EMF performance is low, while FIF achieves higher performance than EMF as the size of FIB increases. EMF can only forward an Interest if an exact match exists on FIB or satisfy an Interest from the CS if its name exactly matches the name of a stored data packet. For FIF though, larger FIB sizes denote a more diverse and complete namespace knowledge by each router, therefore it is more likely to find a fuzzy match that satisfies the selected semantic threshold. 

\subsection {FIF Overhead}

\begin{figure}[t!]
  \centering
  \includegraphics[width=\columnwidth]{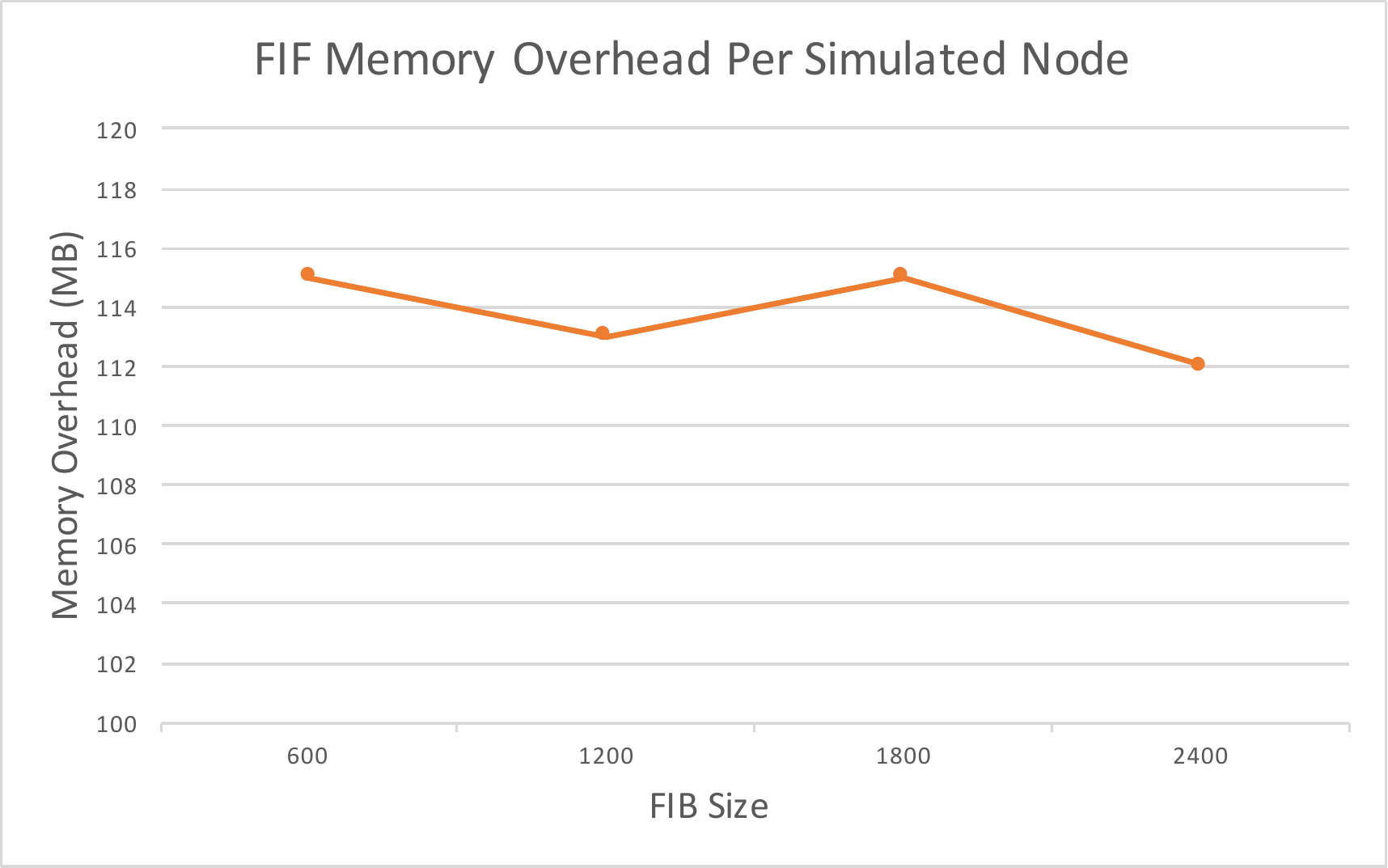}
  \caption{FIF Memory Overhead, number of matches = 50, semantic similarity threshold = 0.4}
  \label{fig:over}
\end{figure}

In Figure~\ref{fig:over}, the FIF memory overhead per simulated node is illustrated. This overhead is estimated by subtracting the overall memory consumption of the EMF-based simulation scenarios from the overall memory consumption of the FIF-based ones and dividing by the number of simulated nodes. The memory overhead per simulated node stays, more or less, constant at about 115MB as we alter the FIB size. 

In terms of time, computing the semantic similarity between 2 words ranges between about 50 - 500 microseconds with an average of 250 microseconds. If a fuzzy match is found by the first hop router of the consumer, the name of the matched FIB entry can be attached to the Interest. Assuming that a routing announcement has been propagated properly all the way from the producer to the consumer, it is likely that the same FIB entry may exist at the next-hop router. By directly forwarding the received Interest based on the attached fuzzy matched FIB entry name, the next-hop forwarder can avoid performing further fuzzy lookups, thus minimizing the imposed time overhead due to FIF.

\subsection {FIF Scalability}

\begin{figure}[t!]
  \centering
  \includegraphics[width=\columnwidth]{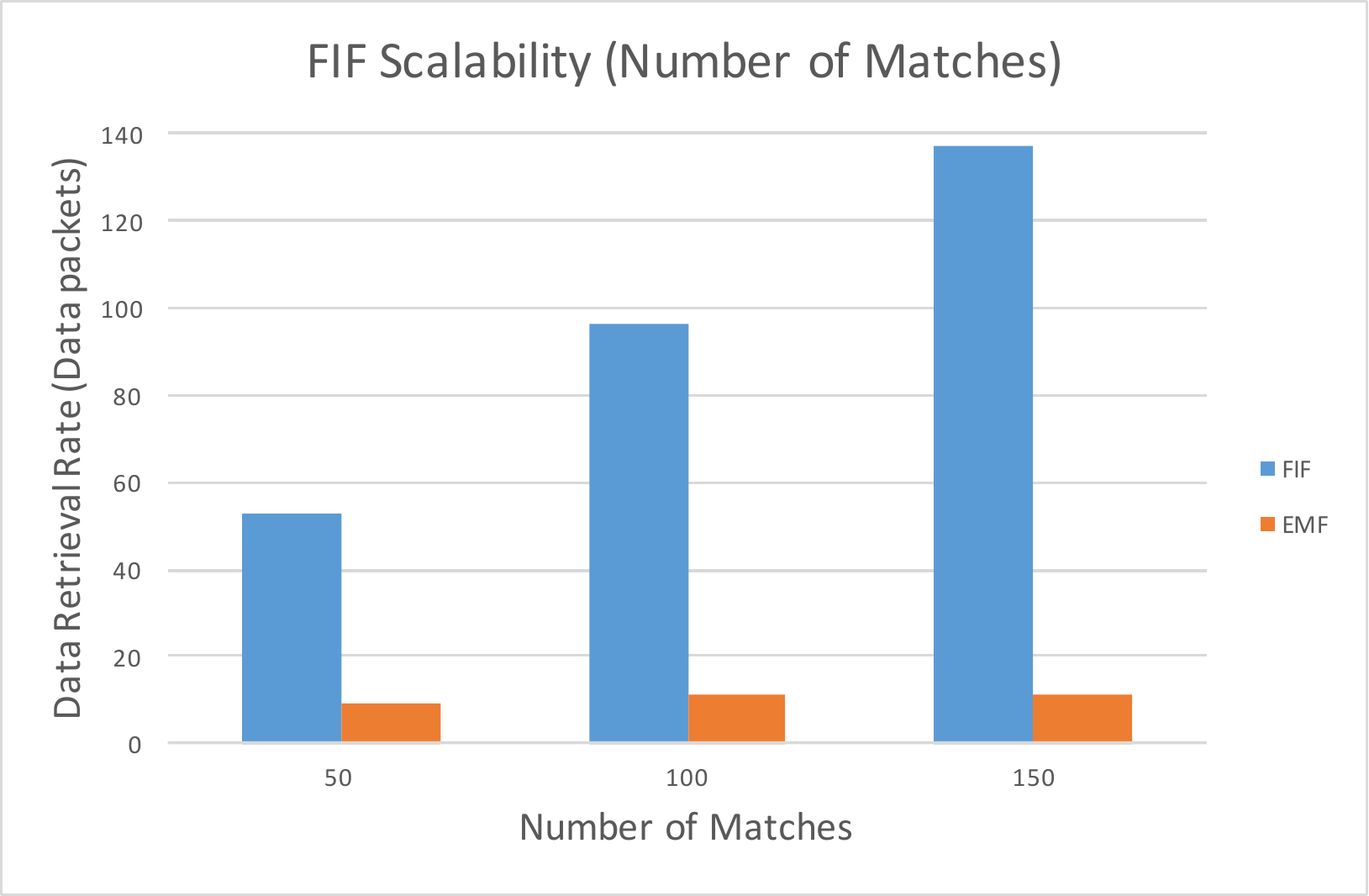}
  \caption{FIF Scalability, varying number of matches, semantic similarity threshold = 0.4, FIB size = 1200}
  \label{fig:scala-matches}
\end{figure}

In Figure~\ref{fig:scala-matches}, we illustrate the data retrieval rate of FIF as we increase the returned number of matches (aggregated number of CS and FIB matches). We observe that the greater the number of matches is, the better the performance of FIF is compared to EMF. However, FIF scales sub-linearly to the number of matches. Compared to EMF, for 50 matches, FIF performs 5.7x better, while for 100 and 150 matches, it performs 8.7x and 12x better respectively.

\begin{figure}[t!]
  \centering
  \includegraphics[width=\columnwidth]{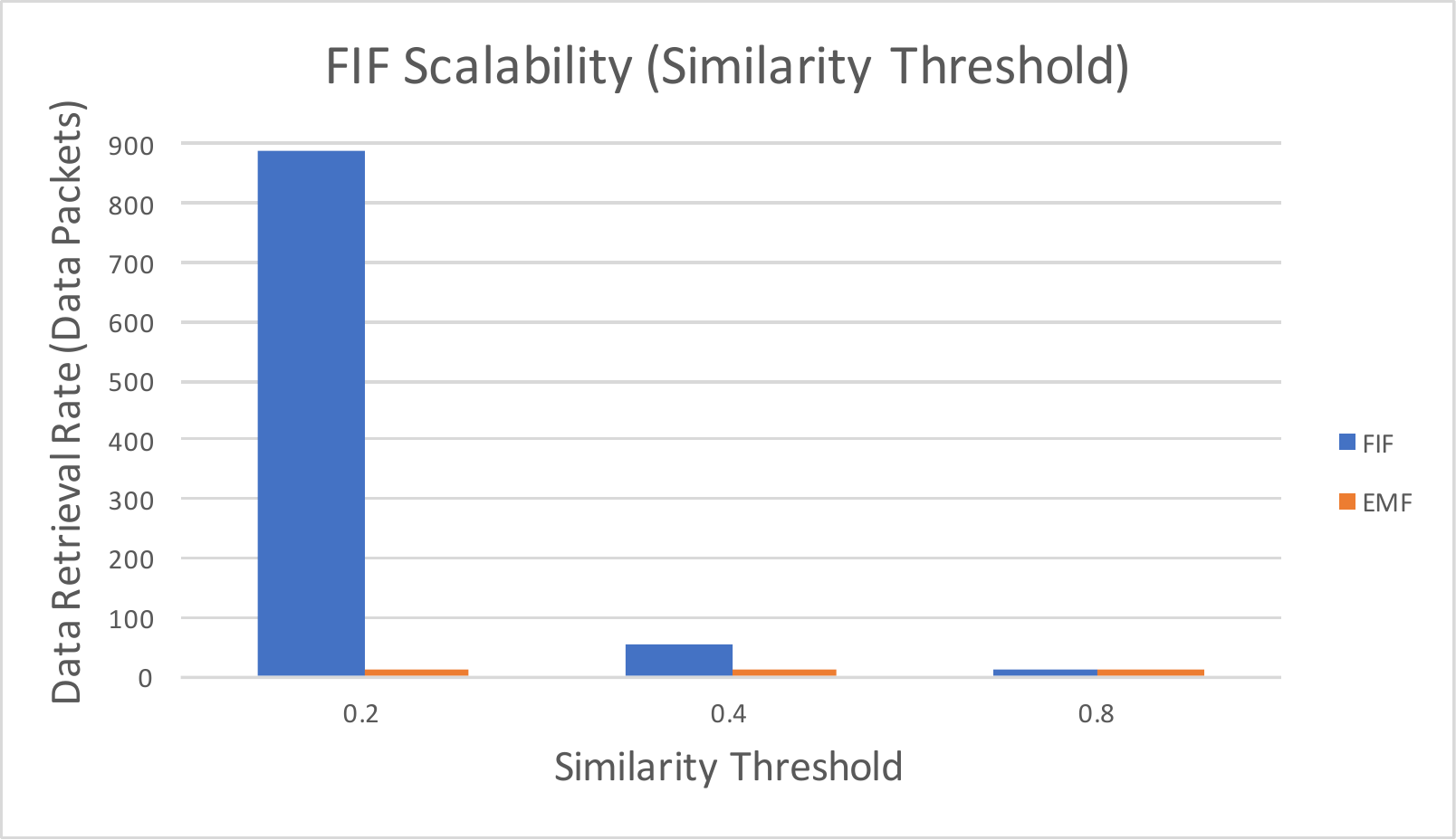}
  \caption{FIF Scalability, varying threshold, number of matches = 50, FIB size = 1200}
  \label{fig:scala-threshold}
\end{figure}

In Figure~\ref{fig:scala-threshold}, we illustrate the data retrieval rate of FIF as we vary the similarity threshold needed to either forward an Interest based on a fuzzy matched FIB entry or satisfy an Interest with a fuzzy matched data packet from the CS. Since the cosine similarity of a large number of words in the training set of the used model is between 0.2-0.3, FIF is able to find a fuzzy match for most of the Interests when the similarity threshold is equal to 0.2. As we increase the similarity threshold, the FIF retrieval rate drops exponentially, since exponentially fewer words in the training set can satisfy the similarity threshold requirement. Compared to EMF, FIF performs 73x, 5.7x and 1.1x better for similarity threshold equal to 0.2, 0.4 and 0.8 respectively.

\subsection {``Forward or Wait'' Performance}

As mentioned in Section~\ref{setup}, in this experiment we deploy one consumer application at node 0 and one at node 3. We aim to evaluate the effect of waiting before forwarding an Interest with an initial fuzzy match or rejecting an Interest that has no initial fuzzy match on the performance of node 3 in the following cases: 1) there is a varying overlap percentage between the set of Interest prefixes expressed by node 0 and node 3, and 2) there is a varying wait time before performing a new fuzzy CS lookup to find a better (in terms of cosine similarity) match than the initial one for node 3. 

\begin{figure}[t!]
  \centering
  \includegraphics[width=\columnwidth]{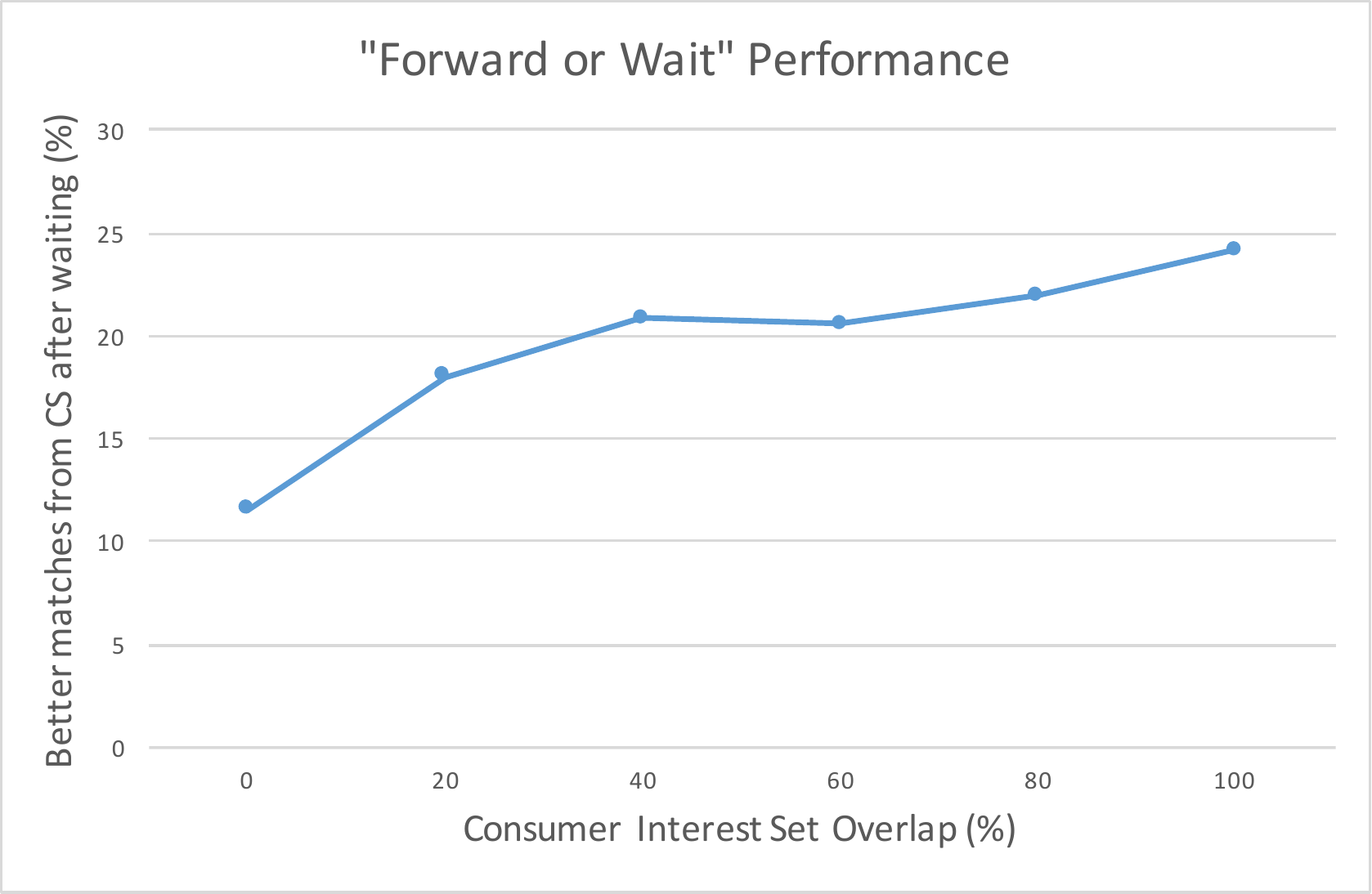}
  \caption{``Forward or Wait'' performance with varying consumer Interest set overlap, wait time = 3xRTT}
  \label{fig:wait-or-fwd-over}
\end{figure}

In Figure~\ref{fig:wait-or-fwd-over}, we present the percentage of Interests at node 3 that resulted in a better fuzzy match found in the CS after waiting as we vary the percentage of the consumer Interest set overlap. The results show that as we increase the overlap of the sets of the Interest prefixes expressed by the consumers, the percentage of better CS fuzzy matches found in the CS after waiting increases for node 3. However, even in the case that the consumers express Interests with exactly the same names, about 25\% of the Interests resulted in better CS matches after waiting. This has to do with the following factors: 1) there is no better match in the CS even after the end of the wait period, 2) due to the limited number of returned matches, there is a better match in the CS, however, it could not be found, and 3) there used to be a better match in the CS, but it happened to be evicted. 

\begin{figure}[t!]
  \centering
  \includegraphics[width=\columnwidth]{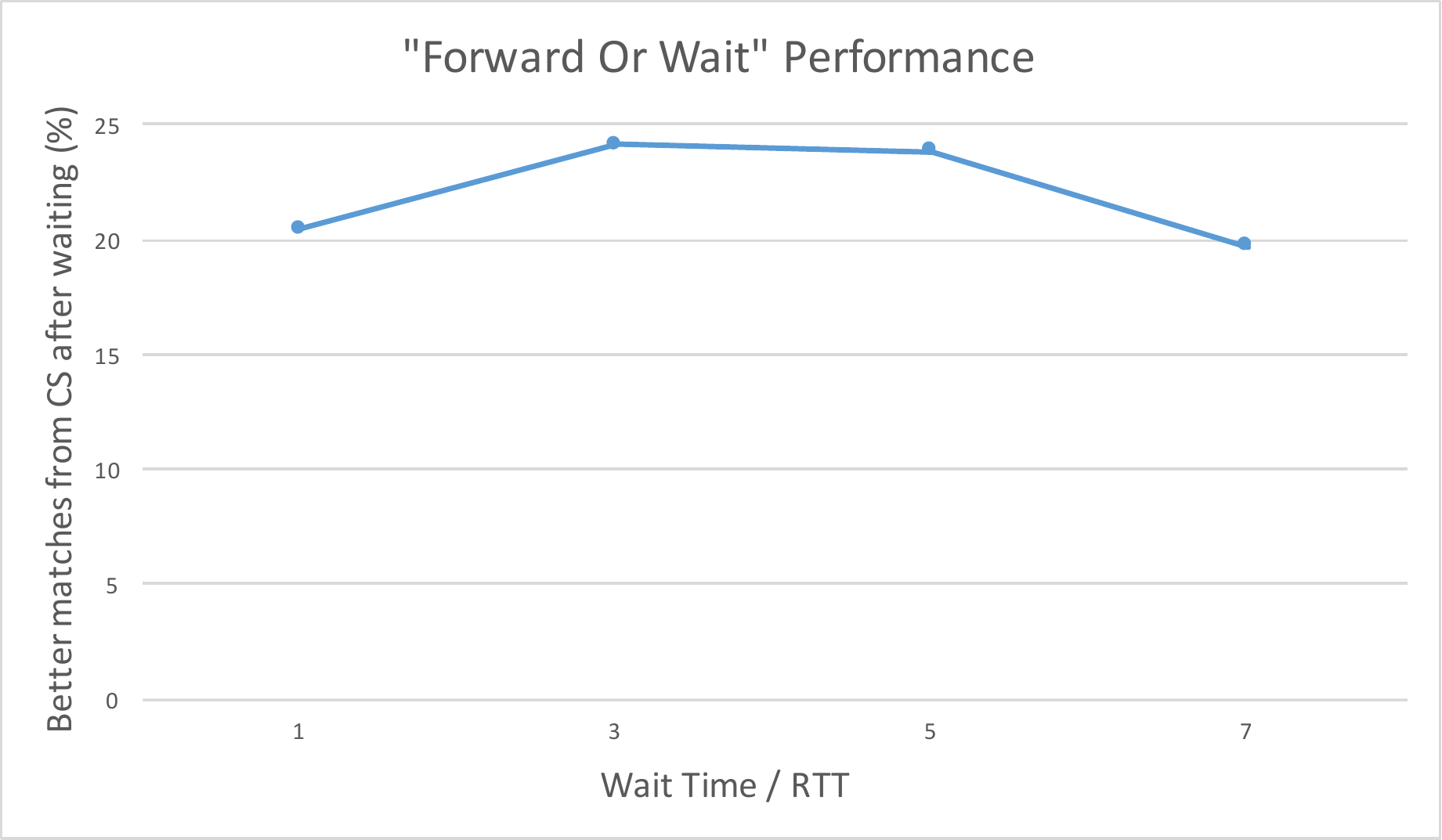}
  \caption{``Forward or Wait'' performance with varying wait time, consumer Interest set overlap = 100\%}
  \label{fig:wait-or-fwd-time}
\end{figure}

In Figure~\ref{fig:wait-or-fwd-time}, we present the percentage of Interests at node 3 that resulted in a better fuzzy match found in the CS after a varying wait time in terms of RTT between node 3 and the producer (node 10). To focus on the effect of the varying wait time on the quality of the fuzzy matches at node 3, the consumers express Interests with the same name. The results show that the highest percentage of better fuzzy CS matches occur for wait times of 3 and 5 RTTs, since there is sufficient time for more diverse data packets to be stored in the CS of node 3. For longer wait times (7xRTT), the number of better fuzzy CS matches drops, since they start getting evicted from the CS. Therefore, either for smaller (1xRTT) or larger wait times (7xRTT), about 20\% of the forwarded Interests result in a better CS match. We should note that the cosine similarity between the better match found in the CS after waiting and the fuzzy component of the Interest name is, on average, 12\% higher than the cosine similarity between the initial match and the fuzzy component of the Interest name.

We conclude that it is hard to specify what would be the optimal wait time. To this end, the approach of using a dynamic timer can be employed. The value of the timer can be adjusted based on the NDN feedback loop created during the exchange of Interests and data. 

\subsection {Application-Observed Retrieval Accuracy}

\begin{figure}[t!]
  \centering
  \includegraphics[width=\columnwidth]{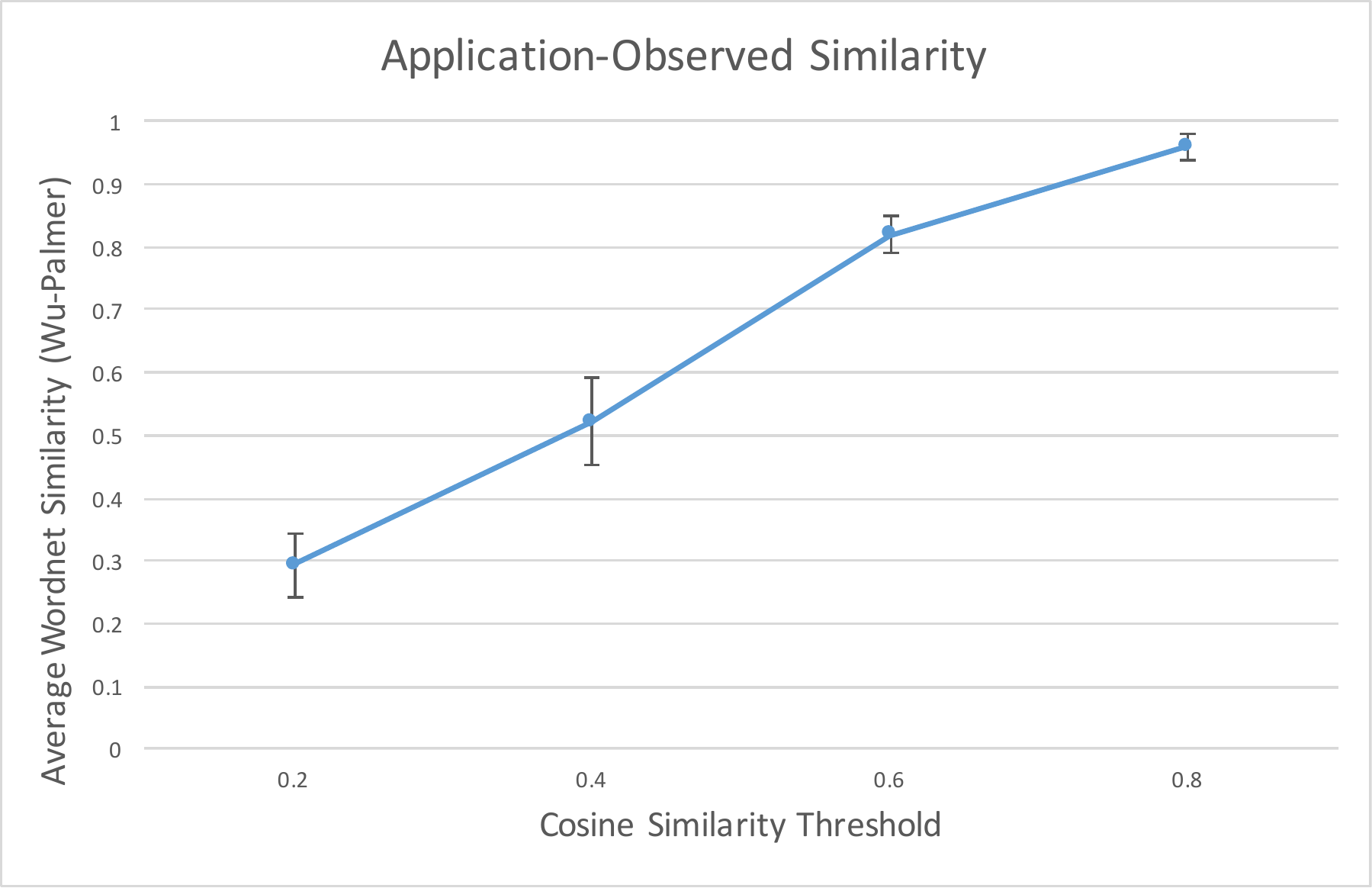}
  \caption{Application-Observed Retrieval Accuracy according to Wordnet}
  \label{fig:appl-obs-acc}
\end{figure}

To estimate the similarity between the retrieved data and the expressed Interests as observed by the consumer application, we collect their names during the simulations and we use the Wordnet lexical database to post-process them after the end of the simulations. In Figure~\ref{fig:appl-obs-acc}, we present the application-observed WUP similarity (accuracy) according to Wordnet for a varying cosine similarity threshold. The results show that the application experiences in general higher similarity rates compared to the in-network similarity threshold. 

The acceptable similarity among the expressed Interests and data packets have to do with the requirements of each application. As a rule of thumb though, it seems that meaningful data (in terms of the application-observed similarity) can be retrieved for in-network cosine similarity threshold close to 0.4.

\section{Identified Challenges}
\label{sec:challenges}

\paragraph{Resource Constraints} The current semantic similarity function, as described in Section \ref{sec:fuzzy-name-matching-lookup} and highlighted in our preliminary evaluations in previous Section, can have a high demand on storage and computational resources (compared to exact matching-based forwarding), especially for nodes in resource-constrained environments. While the quantitative assessment of the approaches and their optimization call for further research efforts (in terms of memory requirements, processing speed, and possibly use of hardware acceleration), 
one way to get around this difficulty is to use default forwarding to direct interests that fail local matching to a more powerful node 
(i.e., semantic oracle) providing the semantic analysis for 
the fuzzy match computation. This removes the computation and storage cost on individual nodes, but still offers the FIF capability if adequate network bandwidth is available. In a bandwidth limited environment, we may consider either to use a fast, but possibly lower performing approach to locally carry out fuzzy matching,
or to let non-fuzzy nodes perform forwarding according to the default routes.

\paragraph{Improving matching through feedback}  When a forwarding node fails fuzzy matching in FIB lookup, another option can be sending a negative acknowledgment (NACK) to the requester, together with the most relevant names/name prefixes.  This NACK can be potentially useful to the requester in two ways: 
a) to inform what contents may be available, and
b) to guide the adjustment of the requested name and try again.
A related challenge is how to incorporate this additional knowledge in the semantic model.  One way could be to add dynamically-adjusted weights (or `penalty') to the semantic distance of those NACK'ed, and another to create a companion model for ``black lists''.  The details of how to do these systematically remain an open issue. 

%

\paragraph{Attack Surface with Fuzzy Matching}  
Although one can stay assured that NDN's data-centric security design should be able to detect all false data, malicious parties may still try to abuse FIF to launch DoS attacks.  Approaches similar to the one described in~\cite{afanasyev2013interest-flooding} can be explored as potential solutions.

%

\section {Conclusion}
\label{sec:conclusion}

In this work, we presented a fuzzy Interest forwarding approach that enables the discovery and retrieval of data using approximate knowledge of the data namespace to augment the NDN forwarding plane. Our simulation study showed that FIF achieves orders of magnitude higher data retrieval rates than EMF at the cost of retrieving data that might not fully match or be necessarily relevant to the data requested by the consumer. FIF scales both in terms of the number of matches and the semantic similarity threshold, however, it does impose considerable overhead on router processing both in terms of time and the memory needed.

\section{Acknowledgments}

This work is partially sponsored by the U.S. Army Research Laboratory and the U.K. Ministry of Defense under Agreement Number W911NF-16-3-0001, and the National Science Foundation under awards CNS-1345318 and CNS-1629922. The views and conclusions contained in this document are those of the authors and should not be interpreted as representing the official policies, either expressed or implied, of the U.S. Army Research Laboratory, the U.S. Government, the U.K. Ministry of Defense or the U.K. Government. The U.S. and U.K. Governments are authorized to reproduce and distribute reprints for Government purposes notwithstanding any copyright notation hereon.


\bibliographystyle{plain}
\bibliography{refs}
\balance

\end{document}